\documentstyle[aps]{revtex}
\twocolumn
\begin{document}
\draft
\title{Giant Spin Relaxation Anisotropy in Zinc-Blende Heterostructures}

\author{N.S.~Averkiev and L.E.~Golub}
\address{A.F.~Ioffe Physico-Technical Institute, Russian Academy
 of Sciences,  194021 St.~Petersburg, Russia}
\maketitle

\begin{abstract}
Spin relaxation in-plane anisotropy is predicted for heterostructures
based on zinc-blende semiconductors. It is shown that it manifests
itself especially brightly if the two spin relaxation mechanisms
(D'yakonov-Perel' and Rashba) are comparable in efficiency. It is
demonstrated that for the quantum well grown along the [001] direction,
the main axes of spin relaxation rate tensor are [110] and
[1$\bar{1}$0].
\end{abstract}

\section{Introduction}

Spin relaxation processes have significant effect in optical and kinetic
properties of semiconductors. They play important role in optical orientation
of electrons and nuclei~\cite{Zakh} and in anomalous magnetoresistance caused
by weak localisation~\cite{Hikami}. Both theoretical calculations and
experimental data analysis have been carried out assuming that one spin
relaxation mechanism
dominates only. Therewith in spite of the strong anisotropy of spin-orbit
scattering, the relaxation times of spin lying in the plane of a
heterostructure with zinc-blende lattice turn out to be independent on
orientation with respect to crystallographic axes.

Real heterostructures differ from investigated ideal objects in that several
spin relaxation mechanisms exist~\cite{Pikus/theor,Knap}.
A spin relaxation mechanism due to only cubic in wave vector terms of
the bulk Hamiltonian was investigated for rectangular quantum wells (QWs)
in Ref.~\cite{DK}. It was noted that even in an asymmetrical GaAs QW, the
efficiency of another mechanism due to linear in two dimensional wave vector
terms is negligibly small. The authors of Ref.~\cite{Knap}
analysing experimental data on anomalous magnetoresistance in InGaAs QWs
demonstrated that the both mechanisms may be comparable in efficiency. But in
Refs.~\cite{Pikus/theor,Knap}, it was mentioned that the both mechanisms are
additive in spin relaxation.

This communication is devoted to an investigation of spin relaxation processes
when several mechanisms of spin-orbit scattering exist.
We show that contributions of these mechanisms interfere and
their simultaneous action leads to the strong anisotropy of spin relaxation even
in the plane of a QW.

\section{Theory}

In zinc-blende semiconductors, spin relaxation of electrons is well known
to be due to spin-orbit splitting of conduction band. In a bulk crystal, the
splitting is cubic in wave vector. In a QW structure, the corresponding
Hamiltonian has to be averaged over the motion along the growth axis.
We consider the QW
grown along $z$-direction parallel to [001] and choose $x$ and $y$
directions coinciding with crystallographic axes. At
relatively small carrier concentrations, one can neglect cubic in 2D wave
vector terms and the Hamiltonian has the form:

\begin{equation}
\label{H1}
H_1 = a_1 (\sigma_x k_x - \sigma_y k_y) \:.
\end{equation}
Here $\sigma_i$ ($i=x,y$) is the Pauli matrix, $k_i$ is the wave vector
component in the plane of the QW and $a_1$ is a constant.

In asymmetrical heterostructures, there is a contribution to the Hamiltonian
which is absent in the bulk~\cite{Rashba}:

\begin{equation}
\label{H2}
H_2 = a_2 (\sigma_x k_y - \sigma_y k_x) \:,
\end{equation}
where $a_2$ is the constant determined by heterointerface properties.

To calculate spin relaxation times, one has to consider the dynamics of
spin density matrix, $\rho$:

\begin{equation}
\label{densmat1}
{\partial \rho \over \partial t} = - {i \over \hbar} \left[ H, \rho
\right] \:, \end{equation}
where the total two dimensional Hamiltonian is:

\begin{equation}
\label{Htotal}
H = {\hbar^2 k^2 \over 2 m} + V + H' \:.
\end{equation}
Here $m$ is an effective electron mass,
$V (x,y)$ is a scattering potential and
\[ H'= H_1 + H_2 \:.\]
We assume that the scattering is elastic and independent on spin indices.

Since $H'$ is a small perturbation, the spin relaxation times turn out to be
much longer than isotropisation times of momentum distribution of electrons.
For this reason it is convenient to represent the density matrix as a
sum~\cite{Pikus/Titkov}:

\[\rho = \overline{\rho} + \rho' \:, \hspace{2cm} \overline{\rho'} = 0
\:, \]
where the bar means averaging over the directions of ${\bbox k}$.
Here $\overline{\rho}$ depends on $\varepsilon = \hbar^2 k^2 / 2 m$ and
the anisotropic part of the density matrix is due to $H'$ only. Hence
$\rho'$ is proportional to $H'$, i.e. $\rho'$ is small in comparison to
$\overline{\rho}$.

Then in the first order in $H'$, the Eq.~(\ref{densmat1}) has the form:

\begin{equation}
\label{densmat'}
{\partial \rho' \over \partial t} = - {i \over \hbar} \left[ H',
\overline{\rho} \right] - \sum\limits_{\bf k'} W_{\bf k k'} \left[
\rho' (\bbox k) - \rho' (\bbox k') \right] \:,
\end{equation}

\begin{equation}
\label{densmat/averaged}
{\partial \overline{\rho} \over \partial t} = - {i \over \hbar}
\overline{ \left[ H', \rho' \right] } \:. \end{equation}
Here $W_{\bf k k'}$ is the probability for scattering from the
potential $V$ from the state with ${\bbox k}$ to the state with ${\bbox
k}'$.

One can see from the Eq.~(\ref{densmat'}) that $\rho'$ relaxes in the time
which is of order of momentum relaxation time, but $\overline{\rho}$ relaxes
in the longer time
which is determined by $H'$ | see~(\ref{densmat/averaged}). At these
long times, the Eq.~(\ref{densmat'})
reduces to a quasistationary one:

\begin{equation}
\label{densmat'2}
\sum\limits_{\bf k'} W_{\bf k k'} \left[ \rho' (\bbox k) - \rho' (\bbox
k') \right] = - {i \over \hbar} \left[ H', \overline{\rho} \right] \:.
\end{equation}

Finding $\rho' (\bbox k)$ from here and substituting it to the
Eq.~(\ref{densmat/averaged}), one can obtain the closed equation for
$\overline{\rho}$:

\begin{equation}
\label{densmat/final}
{\partial \overline{\rho} \over \partial t} = - {1 \over \hbar^2}
\sum\limits_n \tau_n \left[ H'_{- n}, \left[ H'_n, \overline{\rho}
\right] \right] \:.\end{equation}
Here

\begin{equation}
\label{H'/n}
H'_n = \oint \: {d \varphi_{\bf k} \over 2 \pi} \: H'({\bbox k})
\exp (-i n \varphi_{\bf k} ) \:, \end{equation}
where $\varphi_{\bf k}$ is the angle between ${\bbox k}$ and $x$ axis,
and

\begin{equation}
\label{tau/n}
{1 \over \tau_n} = \oint \: d \theta \: W_{\bf k k'} (1 -
\cos n \theta ) \:, \end{equation}
where $\theta = \varphi_{\bf k} - \varphi_{\bf k'}$.
The Eq.~(\ref{densmat/final}) clearly demonstrates that
it is spin-orbit interaction which causes $\overline{\rho}$ relaxation.

After substituting $\overline{\rho}$ in a form

\[ \overline{\rho} = f_0 + {1 \over 2} {\bbox \sigma} \cdot
{\bbox \ae} \:, \]
the Eq.~(\ref{densmat/final}) reduces to following equations:

\begin{equation}
\dot{f_0} (\varepsilon, t) = 0 \:, \end{equation}

\begin{equation}
\dot{\ae}_i (\varepsilon, t) = - \Gamma_{ij} (\varepsilon) \:
\ae_j (\varepsilon, t) \:, \end{equation}
where

\begin{equation}
\label{Gamma}
\Gamma_{ij} =- {1 \over \hbar^2} {\rm Tr} \left\{ \sum\limits_n \tau_n \left[
H'_{- n}, \left[ H'_n, \sigma_j \right] \right] \sigma_i \right\} \:.
\end{equation}

An initial condition may be derived considering the spin dynamics after the
time $\tau_n$, but before the spin relaxation time. In the time $\tau_n$, the
density matrix becomes isotropic but the spin relaxation processes do not start
yet. Therefore~\cite{DP}:

\begin{equation}
\label{f0}
f_0 (\varepsilon) = {1 \over 2} \: [F_+ (\varepsilon) + F_-
(\varepsilon)] \:,\end{equation}

\begin{equation}
\label{kappa0}
{\bbox \ae} (\varepsilon) = {\bbox s} \: [F_+ (\varepsilon) - F_-
(\varepsilon)] \:,\end{equation}
where ${\bbox s}$ is the unit vector along the spin and $F_\pm (\varepsilon)$
are distribution functions of particles with the spin projection on ${\bbox s}$
equal to $\pm 1/2$.

Taking into account that the spin density, ${\bbox S} (t)$, is the average of
${\bbox \ae}$ over $\varepsilon$, one can obtain the kinetic equation for it
at the time longer than $\tau_n$:

\begin{equation}
\label{S0}
\dot{S}_i = - {S_j \over \tau_{ij} } \:, \end{equation}
where the tensor of reciprocal spin relaxation times is:

\begin{equation}
\label{tau}
{1 \over \tau_{ij} } = { \int d \varepsilon \: [F_+ (\varepsilon) -
F_- (\varepsilon)] \Gamma_{ij} (\varepsilon) \over \int d \varepsilon
\: [F_+ (\varepsilon) - F_- (\varepsilon)]} \:. \end{equation}

The Eq.~(\ref{tau}) represents the extension of the results of
Ref.~\cite{DK} to the case of an arbitrary spin-orbit interaction $H'$
and takes into account the anisotropy of scattering.

Substituting $H' = H_1 + H_2$ into~(\ref{Gamma}) and then $\Gamma_{ij}$
into~(\ref{tau}) we have:

\begin{equation}
\label{tauz}
{1 \over \tau_{zz}} = C \: (a_1^2 + a_2^2) \:, 
{1 \over \tau_{zx}} = {1 \over \tau_{zy}} = 0 \:, \end{equation}

\begin{equation}
\label{tauperp}
{1 \over \tau_{xx}} = {1 \over \tau_{yy}} = {C \over 2} \: (a_1^2 +
a_2^2) \:, 
{1 \over \tau_{xy}} = - C \: a_1 \:
a_2 \:, \end{equation}
where

\begin{equation}
\label{C}
C= {8m \over \hbar^4} { \int d \varepsilon \: [F_+ (\varepsilon) - F_-
(\varepsilon)] \tau_1 (\varepsilon) \: \varepsilon \over \int d
\varepsilon \: [F_+ (\varepsilon) - F_- (\varepsilon)]} \:. \end{equation}

Equations~(\ref{tauz} | \ref{C}) generalize the
results~\cite{Pikus/theor,Knap,DK}
for the case of two spin relaxation mechanisms.

\section{Discussion}

It follows from the Eq.~(\ref{tauperp}) that because of the two mechanisms
($a_1 \cdot a_2 \neq 0$), the spin relaxation in the plane of the QW becomes
anisotropic. It should be emphasized that if there is only one mechanism
($a_1 \cdot a_2 = 0$), then the spin relaxation is isotropic in spite of
the cubic symmetry of the Hamiltonian $H_1$ or $H_2$. Thus the cubic
anisotropy of the real QW structure manifests itself due to the interference of
two spin relaxation mechanisms.

The system~(\ref{S0}) may be rewritten as follows:

\begin{equation}
\label{S+-}
\dot{S}_x \pm \dot{S}_y = - {S_x \pm S_y \over \tau_\pm} \:, \end{equation}
where

\begin{equation}
\label{taupm}
{1 \over \tau_{\pm} } = {C \over 2} (a_1 \pm a_2)^2 \:. \end{equation}
The times $\tau_+$ and $\tau_-$ describe the relaxation of the spin oriented
along the directions [110] and [1$\bar{1}$0] respectively.

The most bright manifestation of spin relaxation anisotropy occurs if
$a_1 = \pm a_2$. In this case, one of the times $\tau_+$ or $\tau_-$
becomes infinite. Therewith the other is equal to $\tau_{zz}$.

Efficiency of the mechanisms depends on both the material and the shape of
the QW. It was shown that in a rectangular GaAs/AlGaAs QW, the
mechanism~(\ref{H1}) dominates~\cite{Klitzing}, and in an asymmetrical
InGaAs/AlAs QWs the mechanism~(\ref{H2}) is the most important~\cite{Savel'ev}
or they are comparable~\cite{Knap}.

It is clear that if both $a_1$ and $a_2$ are not equal to zero, then
the spin sublevels split. In the work~\cite{Wissinger} it is shown that
$a_1$ and $a_2$ may be comparable in magnitude and, hence, the spin
splitting is strongly anisotropic.

The spin relaxation anisotropy results from the initial
$T_d$ symmetry of the zinc-blende semiconductor. For this reason, the similar
effect can take place in a strained bulk crystal. The corresponding Hamiltonian
linear in 3D wave vector, ${\bbox k}$, and components of an elastic
strain tensor, $u_{ij}$, has the form:

\begin{eqnarray}
\label{Hdef}
H' (u) = A_1 \: u_{ii} \: (\sigma_{i+1} k_{i+1} - \sigma_{i+2}
k_{i+2}) \\ + A_2 \: u_{ij} \: (\sigma_i k_j - \sigma_j k_i) \:. \nonumber
\end{eqnarray}
Here $i,j = x,y,z$, $i+3 \to i$, $A_1$ and $A_2$ are constants. Doing
calculations for $1/\tau_{ij}$ in a way similar to~(\ref{tau}), one can
obtain three different spin relaxation times. It can be shown that the
maximum anisotropy may be achieved if

\begin{equation}
\label{u}
A_1 u_{xx} = A_1 u_{yy} = - A_1 u_{zz}/2 = A_2
u_{xy} / 3
\end{equation}
with the rest of $u_{ij} =0$. Therewith two spin relaxation times
are equal to each other and the third is infinite.
Note that the tensor $u_{ij}$ determined by~(\ref{u}) may be obtained by
applying two uni-axial strains along the axes [001] and [110] and they
are not restricted to uni-axial strain along any axes.

\section{Conclusion}

The possibility for spin relaxation suppression was noted in
Ref.~\cite{DK} for a QW grown along [110] direction when the spin is oriented
along the same axis. The present work shows that the spin relaxation
rate also decreases for [110] direction, but in a QW grown in the symmetrical
direction [001]. Therefore this decrease takes place for the
spin lying in the plane of the QW.

Analysing weak localisation effect, the authors of
Refs.~\cite{Pikus/theor,Knap}
showed that the mechanisms~(\ref{H1}) and~(\ref{H2})
suppress each other in anomalous magnetoresistance, but
they are additive in spin relaxation. The
present analysis shows that the suppression occurs in the spin relaxation
also. Besides, we have found that spin relaxation is anisotropic even in
the plane of the QW.

\mbox{}\\ We thank U.~R\"ossler, M.I.~Dyakonov and A.N.~Titkov for
fruitful discussions.

\mbox{}\\ This work was financially supported by
Russian Foundation for Basic Research (grant 98-02-18424) and programm
``Physics of Solid State Nanostructures'' (grant 97-1035).

\end{document}